István Horváth
# The Great Debate

A hundred years ago (1920) in the auditorium of the Smithsonian Institution's U.S. National Museum there were two lectures under the auspices of the George Ellery Hale Lecture series, what has come to be called the "Great Debate". In the debate, Harlow Shapley and Heber Curtis argued over the "Scale of the Universe". Curtis argued that the Universe is composed of many galaxies like our own and they are relatively small. Shapley argued that the Universe was composed of only one big Galaxy. In Shapley's model, our Sun was far from the center of this great island Universe. This article was published in the 2020 Astronomical Yearbook (Hungarian, MCSE).

Horváth István
A Nagy (Shapley–Curtis) Vita

*„Teremté tehát Isten a két nagy világító test ... és a csillagokat. És helyezteté Isten azokat az ég mennyezetére."* Genezis 1, 16-17

Az 1920-as években a világról alkotott képünk jelentős változáson ment keresztül.[1] 1930-ra már nem csak galaxisról, hanem galaxisokról (sőt galaxis halmazokról) beszéltünk, és lényegében most is az akkor kialakult kép van előttünk, ha a Világegyetem felépítésére gondolunk. Az évtized csillagászati felfedezései előtt, foglalta össze a (USA) Nemzeti Tudományos Akadémia (National Academy of Sciences, NAS) 1920-as Washington-i ülésén két előadás, ami később a nagy vita (great debate) nevet kapta, az időszerű problémákat, kérdéseket, és csokorba szedték az addigi ismeretanyagokat. Ezen vita 100. évfordulója van 2020-ban. Ennek emlékezetére íródott jelen cikkünk.

**Már a régi görögök is**

Világképünk, a világról (mint írtuk, kozmoszról) alkotott elgondolásunk többször változott, közelítve a ma használatos elképzelésekhez.[2] Az ember évezredek óta próbálja megérteni a körülötte lévő világot. A régi kor emberének a világot a körülötte levő történések jelentették. A Föld legyen lapos vagy gömbölyű, életünk színtere, az ókori civilizációkban a teljes világot jelentette. A világ keletkezéséről alkotott képét a legtöbb nép teremtésmítoszban meséli el, melyben az Isten vagy istenek teremtik a világot. A világ akkoriban csak a Föld felszínét az ember életének terét jelentette. Egészen a XVI. század végéig ez csak a hét égitesttel (Nap, Hold és az öt bolygó) és a csillagok kristályszférájával egészült ki.

A világot már a kezdeti gondolkodók égi és földi részre osztották. A kettő egyesítéséhez a modern fizika megszületésére kellett várni egészen a XVII. századig. Addig viszont jórészt különböző törvények által irányítottnak tekintették őket. De már a görögöknél is akadtak kivételek, akik az égi objektumok helyét, mozgását nem istenek behatásával magyarázták. Ilyen volt legkorábban Anaximandrosz (Kr. e. 610-547) a Krisztus előtti hatodik században (Crescenzo, 2000). A földi világ bonyolult ugyan, de a Föld alakját nem csak a görögök ismerték. Már a görögök előtt a babiloniakat is érdekelte az égi jelenségek vizsgálata. Az ókori nézetekben azonos, hogy a csillagok egy közös szférán nyugszanak, hiszen egymáshoz képest (látszólag) nem változik helyzetük.

---

[1] Világon itt a fizikai világot értjük, azaz a Világmindenséget (Világegyetemet), amit a görög kozmosz (κόσμος) szóval is szokás megnevezni, nagyon helyesen, hiszen jelentése világ (esetleg rend, így az ókorban rendezett világ).

[2] Jelen cikkben csak a fizikai világról alkotott tudományos elképzeléseket tárgyaljuk.

Az égi mozgások megfigyelése és előrejelzése (például csillagjóslás) nagy fontossággal bírt. Hipparkhosz (Kr. e. 190-120) és elődei (például Eudoxosz (Kr. e. 390-337), Arisztotelész (Kr. e. 384-322), Apollóniosz (Kr. e. 265-190)) eredményeire és adataira támaszkodva Klaudiosz Ptolemaiosz (90-168) a Krisztus utáni második században írta, az arab fordítás után Almagesztnek nevezett művét. Ebben a gömb alakú Földet tekinti a mindenség középpontjának, mely körül kering a hét égitest és a csillagok szférája.[3] Ptolemaiosz modellje több mint ezer évig lehetővé tette a bolygók égi helyzetének előre történő kiszámítását. Nem véletlenül ezt a világképet ma is ptolemaioszi világképnek nevezzük. Számoszi Arisztarkhosz (Kr. e. 310-230) már a Krisztus előtti harmadik században úgy gondolta, hogy a Föld kering a Nap körül. Ezt a nézetét a középkorban is ismerték, és Kopernikusz (1473-1543) is hallott róla itáliai egyetemi tanulmányai során (Koestler, 2007). Kopernikusz érdeme, hogy kidolgozta a Nap-középpontú rendszer matematikáját. Sajnálatos módon ez nem volt egyszerűbb, mint Ptolemaioszé, sőt mint Arthur Koestler (1905-1983) ezt megállapítja (Koestler, 2007) a kopernikuszi rendszer több epiciklust tartalmazott, mint a ptolemaioszi.[4]

**A modern tudományok születése**

Tycho Brahe (1546-1601) méréseit felhasználva Kepler (1571-1630) jön rá a XVII. század elején, hogy a bolygók mégse körpályán mozognak. Galileo Galilei (1564-1642) és mások távcsövekkel tesznek fontos felfedezéseket; a Vénusz fázisai, a Jupiter négy holdja (melyek bizonyosan nem a Föld körül keringenek), a Hold hegyei stb. A század végére megszületik a fizika (Hooke (1635-1703), Huygens (1629-1695), Newton (1642-1727) stb.). Megszületnek az alapvető fizikai fogalmak (erő, impulzus stb.) és megalkotják az ezek leírásához szükséges matematikai apparátust (differenciál- és integrálszámítás, differenciálegyenletek).

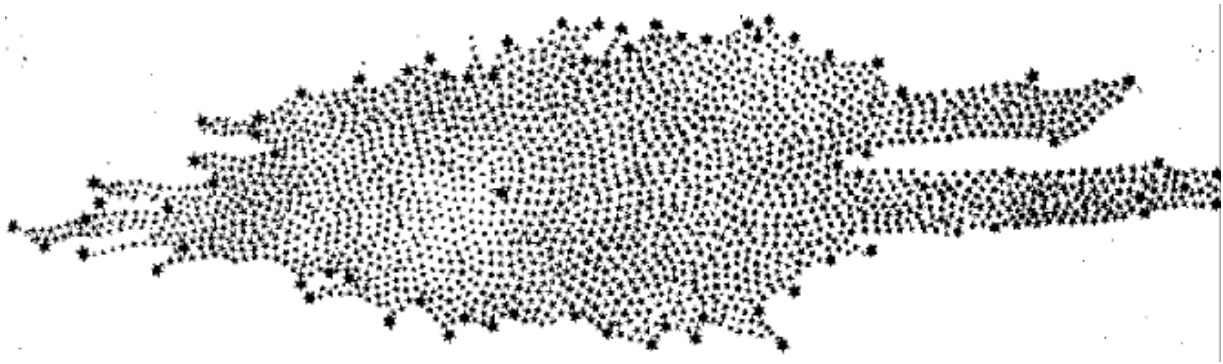

**1. ábra.** Herschel világa, a csillagszámlálásai alapján (Herschel, 1785).

---

[3] Megjegyezzük, hogy az Almageszt tartalmazza Hipparkhosz 850 csillagpozíciójának adatait is. Ez azért fontos, mert Hipparkhosz eredeti munkája nem maradt ránk.
[4] Kopernikusz megközelítése is nagyon hasonló volt Ptolemaioszéhoz, ezért mint azt Koestler is írja, Kopernikusz, műve által az ókori szerzők közé sorolandó. De lényeges, hogy e mű által a Nap-középpontú rendszer gondolata terjedhetett.

A XVIII. századi világkép tehát, a végtelen űr, benne az egymástól távol levő csillagoknak nevezett forró gázgömbökkel. Vannak még nem csillagszerű, hanem kiterjedt világító ködök, és sötét fényelnyelő ködök is. Lényeges módosítást jelentett ezen a képen, hogy William Herschel (1738-1822) csillagszámlálással bebizonyította (Herschel, 1785), hogy egy lapult milliárdnyi csillagból álló korongban helyezkednek el a világunk említett objektumai (1. ábra).[5] A csillagokkal egyenletesen betöltött világűr elképzelést felváltotta egy nagy csillagváros képe. Kérdés volt még hogyan keletkeztek a csillagok, és hogyan keletkezett a csillagváros, illetve maga az egész világ. Vagy a világunk már végtelen ideje létezne?

A XIX. század első felében a legtöbb csillagász a világító ködöket a Tejútrendszer részének tekintette. Tehát világképük szerint a világ sok milliárd csillagból áll ugyan, de ezek egy nagy csillagvárosban a Galaxisban találhatóak. Hogy mi van a Galaktikán kívül, azt nem tudjuk, de valószínűleg semmi. Immanuel Kant (1724-1804) volt az (Thomas Wright (1711-1786) angol földmérő mérnök könyve alapján), aki azt gondolta, hogy a spirális alakú ködök talán ugyanolyan csillagvárosok, mint a mi Tejútrendszerünk (Ferris, 1985). A könyvecske, amiben ezt közzétette, nem volt lényeges hatással tudósvilágra.[6] Annál inkább Pierre-Simon Laplace (1749-1827) elképzelése, aki csillagok keletkezésének helyének gondolta a spirál-ködöket (Laplace, 1796). Így egészen az 1860-as évek végéig szinte mindenki ezt fogadta el. Később a híres spektroszkópus Huggins (1824-1910) megvizsgálta a ködök színképét, és azt találta, hogy a spirálködöké a csillagszínképekre hasonlít, míg sok más ködé a gázok színképére. Ez jó érv volt a spirálködök csillagváros elképzelés mellett. 1885-ben az Androméda-ködben megjelent egy új csillag. Ma már tudjuk, hogy ez egy szupernóva volt. De a XIX. század végén még azt se tudták hogyan termelnek energiát a csillagok. Nem ismerték az elemi részecskéket, nem ismerték az atommagot, az atommagok szerkezetét, hovatovább még azt gondolták, hogy az atomok oszthatatlanok (egyesek az atomok létét is kétségbe vonták). Nem csoda hát, hogy senki se gondolt arra, hogy léteznek olyan csillagok (mint említettük a szupernóvák) melyek heteken keresztül képesek túlragyogni, akár a galaxisukat is.

**A csillagászat tovább fejlődik, fényképezés, spektroszkópia**

Mint említettük, csillagokon és bolygókon kívül kiterjedt objektumokat is megfigyeltek a távcsöves megfigyelők. Ezek egy része csillagokra volt bontható, például a nyílt- vagy a gömbhalmazok. A nem csillagokra bontható kiterjedt alakzatok lehettek fénylő ködök, vagy egyesek szerint nagyon távoli és halvány csillagokból álló csoportosulások. Ezeket az ún. mélyég objektumokat először Charles Messier (1730-1817) listázta 1771 és 1781 között, eredetileg 103 égi ködöt[7], melyet később 110-re bővítettek. William Herschel 1786-ban kiadott első katalógusa 1000 mély-ég objektumot tartalmazott, melyhez Herschel később még ezret hozzátett, majd újabb 500-at. Fia John Herschel (1792-1871) 1864-ben kiadott katalógusában már 5079 szerepelt, ennek neve General Catalogue of Nebulae and Clusters of Stars volt. John Louis Emil Dreyer (1852-1926) katalógusa már New General Catalogue of Nebulae and Clusters of Stars néven, rövidebben New General Catalogue (innen a számunkra ismerős NGC rövidítés) jelent meg 7840 tétellel 1888-ban. Ezeket a fénylő égi ködöket, már maga Herschel is hét kategóriára osztotta. Történetünk szempontjából azok az érdekesek, amelyekben nagy távcsövekben se lehetett csillagokat felfedezni.

---

[5] Herschel eredetileg mindegy 100 millióra becsülte a csillagok számát (Asimov, 1965).
[6] A könyv kiadója tönkrement, így a könyv raktáron maradt (Whitney, 1978).
[7] Az 1784-ben közzétett listát Messier és Pierre Mechain (1744-1804) állította össze (Whitney, 1978).

A távcsöves felfedezéseket jelentős újítások/felfedezések is segítették. A fontosabbak közé tartozik a fényképezés és a színképelemzés. Az első csillagászati felvételt John William Draper (1811-1882) készítette a Holdról 1840-ben. Newton óta használták a prizmát a fény színekre bontására. 1802-ben William Hyde Wollaston (1766-1828) volt az első, aki a Nap fényében sötét vonalakat látott. Wollaston azt hitte ezek a színek természetes határai. Ezt az elképzelést 1815-ben Joseph Ritter von Fraunhofer (1787-1826) megcáfolta. Fraunhofer volt, aki a prizma helyett diffrakciós rácsot használt a színképelemzéshez. Gustav Kirchhoff (1824-1887) és Robert Bunsen (1811-1899) 1859-ben felfedezték, hogy a gázok vonalas színképéből következtetni lehet anyagi minőségükre. Ennek alapján az égitesteket alkotó elemekre is lehet következtetni. Kirchhoff ezt a Napra is alkalmazta. Ekkor született meg az asztrofizika. Az első felvételt csillag csillagszínképről (Szíriusz, Capella) 1863-ban William Allen Miller (1817-1870) és William Huggins (1824-1910) készítette. Az első mély-ég objektumról készített felvételt (Orion-köd) Henry Draper (1837-1882, John William fia) készítette 1880-ban. Eközben a laboratóriumi spektroszkópia is fejlődött. Megállapították, hogy hevített gázok színképében fényes vonalak vannak, azaz csak nagyon kevés frekvencián történik sugárzás. Folytonos színképet gázon átengedve a folytonos színképből a gáz éppen a rá jellemző frekvenciákon elnyeli a sugárzást. A csillagok spektruma tipikusan ilyen. A jelenség fizikáját először majd csak a Bohr féle atommodell segítségével értjük meg 1913-ban[8], de a jelenség már a XIX. században is ismert volt.

Huggins több mély-ég objektumnak (égi ködnek) is vizsgálta a színképét. Voltak közöttük emissziós színképek (mint említettük, ezeknél csak egyes frekvenciákon tapasztalunk sugárzást), ezeket a ködöket helyesen gáz-ködöknek gondolták. Voltak azonban folytonos színképpel rendelkező ködök, például a spirál-ködök, melyeket ma galaxisoknak gondolunk. Ha nem túlságosan fényes csillagok közelében van a csillagközi anyag, akkor nem tudja a csillagfény gerjeszteni, de a köd esetleg ekkor is látható, mert a porról a csillagfény visszaverődhet. Ez esetben a köd fénye hasonló lesz a (környező) csillagokéhoz. Huggins spektroszkópiai mérései döntőek lehettek volna a spirálködök mibenlétével kapcsolatban, vajon Kantnak vagy Laplacenak van-e igaza. Mint említettük Kant elmélete, hogy a spirál-ködök csillagvárosok, észrevétlen maradt. Laplace elképzelése, hogy ezek születőben lévő csillagok, illetve bolygórendszerek terjedt el a tudósok között.

1885-ben Ernst Hartwig (1851-1923) egy új csillagot vett észre az Andromeda-ködben, mely a felfedezést követő napon egészen 5,85 magnitúdóig fényesedett. A kétféle magyarázat egyike, a következő lett volna, egy csillagvárosban felfényesedik egy csillag, mely milliónyi, ha nem milliárdnyi csillag fényességével ragyog. Ma már tudjuk, hogy ez a helyes magyarázat, de 1885-ben ezt csak kevesen gondolták. A másik magyarázat hihetőbb volt, hogy a spirális ködben, amely egy összecsomósódó forgó gázköd, felfénylik egy születőben levő csillag. Kérdés nem lett volna, hogy a magyarázat helyes, ha a csillag évekig, évtizedekig világít, de már fél év múlva is csak 14 magnitúdó fényes volt. Nem baj gondolták a csillagászok a születő csillag először csak felsír, majd mikor stabilizálódik újra világítani fog.

**A csillagászati távolságmeghatározás fejlődése a századfordulón**

Most pedig ismerkedjünk meg Edward Charles Pickeringgel (1846-1919) és az ő számolóival. Pickering 1876-tól haláláig a Harvard Egyetemen dolgozott. Testvérével Williammal (1858-1938) csillagvizsgálót épített az arizonai Flagstaffban és a perui Arequipában. Az itt készült felvételek alapján a csillagok osztályozásával foglalkozott. Összesen 250 ezer felvételt és 225 ezer csillagszínképet készített, melyek elemzésére közel negyven nőt alkalmazott „számolóként" (computer). Ezek közé tartozott Annie Jump Cannon (1863-1941), aki négy egyetemnek is tiszteletbeli doktora lett (az első nő az oxfordin), és Henrietta Swan Leavitt (1868-1921), aki 1895-ben került a Harvardra, ahol azt a feladatot kapta, hogy keressen a fényképeken változócsillagokat. Leavitt több mint 2400 cefeida típusú változócsillagot fedezett fel, és még legalább ugyanennyit vizsgált meg. Egy idő után összefüggést talált a változócsillagok fényessége és változási periódusa között. Ezt azért tudta megtenni, mert a csillagok többsége a Nagy és Kis Magellán-felhő nevű csillagcsoportosulásban volt található. Leavitt 1908-ban publikálta, hogy a fényesebb cefeidáknak hosszabb a periódusuk (Leavitt, 1908), majd 1912-ben kimutatta, hogy a Kis Magellán-Felhő 25 cefeidája esetén e két mennyiség között matematikai összefüggés állítható fel (Leavitt és Pickering, 1912). Ennek segítségével lehetőség nyílt egy jó távolságmeghatározási módszerre. Ennek kalibrálásához meg kellett állapítani legalább egy, de inkább több (közeli) cefeida távolságát.

---

[8] Niels Bohr (1885-1962) a róla elnevezett atommodell felállításáért 1922-ben fizikai Nobel-díjat kapott.

Szintén 1912-höz kapcsolódik egy másik fontos eredmény. Vesto Melvin Slipher (1875-1969) megmérte néhány spirál-ködnek a spektrumát. Slipher a csillagászati spektroszkópia egyik úttörője volt. Egész életében az arizonai Lowell Obszervatóriumban dolgozott, ahova maga Percival Lowell (1855-1916) vette fel 1901-ben, majd nevezte ki megbízott igazgatónak 1916-ban (1929-ben ő alkalmazta Clyde Tombaught (1906-1997) a Plútó későbbi felfedezőjét). Slipher Lowell utasítására kezdte el a spirálködök spektrumának vizsgálatát. 1912 és 1917 között 25 spirális ködnek mérte meg a radiális sebességét (Kun, 2012). 1925-re ez a szám már 45-re növekedett. Átlagos radiális sebességnek a megszokott többszörösét kapta, nagyjából 650 km/s-ot. De mért 1600 km/s értéket is (Ferris, 1985). Néhánytól eltekintve mind tőlünk távolodott. Úgy látszott, hogy ha ezek a Tejútrendszerben is vannak, nincsenek gravitációsan kötve. Ebben az időben (1910-es évek eleje) Ejnar Hertzsprung (1873-1967) és Henry Norris Russell (1877-1957) egymástól függetlenül felfedezte a csillagok színe és fényessége közötti összefüggést (Gribbin, 2004). A nevük után HRD (Hertzsprung-Russell diagram) nevet viselő diagramm látható a 2. ábrán.

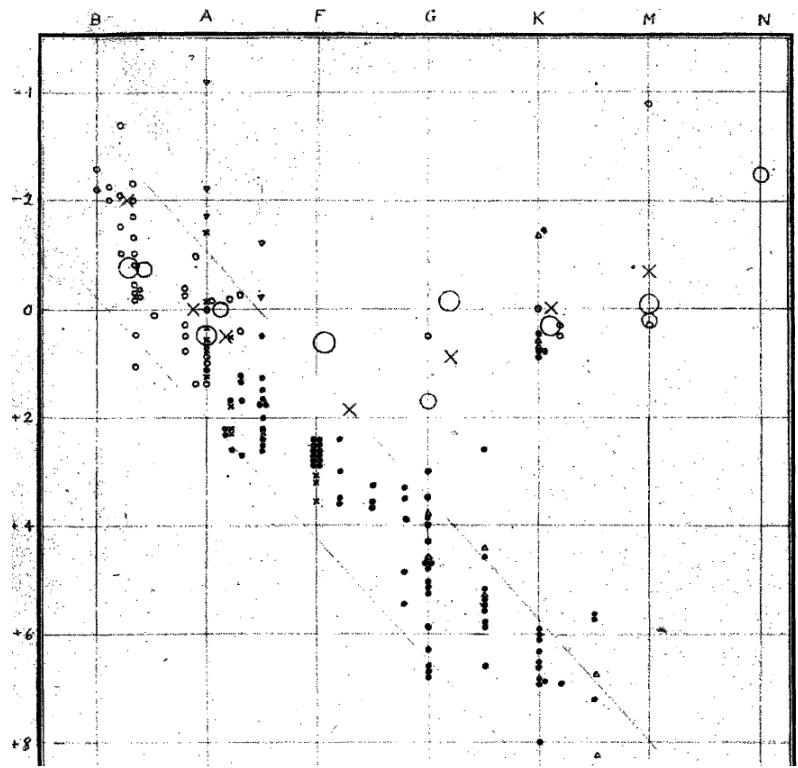

**2. ábra.** A szín-fényesség diagram, Russell eredeti cikkéből (Russell, 1914).

**A vita két előadója**

Ilyen ismeretek után köszöntött be az 1920-as év, melyben a nagy vitát rendezték. A javaslattevő 1919 végén George Ellery Hale (1868-1938) a nagy távcsőépíttető volt.[9] Hale apja William Ellery Hale (aki felvonók gyártásából gazdagodott meg) tiszteletére felajánlást tett a Nemzeti Tudományos Akadémián tartandó előadásokra. Hale javasolta, hogy az 1920-as Washington-i ülésen két előadás hangozzék el vita formájában. Hale két témát javasolt, a sziget csillagvilágok témáját, illetve a relativitáselméletet. Charles Greeley Abbottal (1872-1973) az akadémia titkárával (home secretary) történt egyeztetés után a Világegyetem mérete (The Scale of the Universe) címben maradtak.

A téma időszerű volt, hiszen Shapley nemrég közölte elképzelését arról, hogy a Tejútrendszer szerinte sokkal nagyobb, mint addig gondolták, illetve a Naprendszer távol fekszik annak közepétől. Harlow Shapley a Missouri-beli Nashville-ben született 1885-ben. Állítása szerint véletlenen múlt, hogy csillagász lett belőle.[10] Doktori dolgozatát Henry Norris Russelnél (1877-1957) írta Princetonban fedési változók témában 1914-ben. Ez után Hale meghívására a kaliforniai Mount Wilson Obszervatóriumba ment kutatni. Kutatási területe a gömbhalmazok voltak. A gömbhalmazok cefeida változói megfigyelésével megállapította a közeliek távolságát. Ezekben a cefeidák fényességét összehasonlította a vörös óriások és a szuperóriások fényességével, ezáltal megállapítva azok abszolút magnitúdóját, mely felhasználásával a távoli gömbhalmazok távolságát becsülte meg. A csillagközi anyag (por, gáz) elnyelése még nem volt közismert, ezért csakúgy, mint a többi csillagász, Shapley sem vette figyelembe ezt a hatást. Az akkor elterjedt nézet szerint a Nap a Tejútrendszer közepén foglal helyet, melynek mérete 7-15 ezer fényév.[11] Shapley mérései szerint csillagvárosunk legalább hússzor ekkora[12] és a Naprendszer nem a központban, hanem attól lényegesen távol mindegy 50 ezer fényévre található (jelenlegi elfogadott érték kb. 26500 fényév). Ezen eredményeket Shapley teljesen egyedül érte el, és közölte a mindenki által elfogadott nézetekkel szemben. Ezért őt hívták meg az egyik előadónak. Még a vita évében meghívták a Harvard csillagvizsgáló igazgatójának, melyet el is fogadott. Shapley 1972-ben halt meg Boulderben.

Másik előadónak eredetileg William Wallace Campbellt (1862-1938) a Lick csillagvizsgáló igazgatóját akarták meghívni, de végül is beosztottja Heber Doust Curtis kapta a felkérést. Curtis Michigan államban született 1872-ben. Iskoláit Detroitban végezte, majd diplomáját a University of Michigan-en szerezte. Ez után klasszikus nyelveket (latin, görög) tanított Kaliforniában. 1900-ban önkéntesként részt vett a Lick Obszervatórium napfogyatkozás expedíciójában. Majd Vanderbilt Ösztöndíjjal került a Virginiai Egyetemre, ahol 1902-ben csillagászatból doktorált. Ez után 1920-ig a Lick Csillagvizsgáló munkatársa volt, ahol fő feladata a ködök megfigyelése volt. Curtis akárcsak Shapley munkahelyet váltott 1920-ban, az Allegheny Csillagvizsgáló igazgatója lett. Curtis 1942-ben halt meg Ann Arborban.

---

[9] Hale építette a világ legnagyobb lencsés távcsövét (Yerkes Obszervatórium 102 cm), a Wilson hegyi 1,5 és 2,5 méteres tükrös távcsöveket (Hale távcső 60 inch, Hooker távcső 100 inch), valamint az 5 méteres (200 inch) tükröset a Palomar hegyi csillagvizsgálóban, melyet ugyancsak róla neveztek el.

[10] Újságíró szakra szeretett volna beiratkozni, de abban az évben az nem indult. A szakok névsorában első archeológia szót ki se tudta mondani, ezért a következőre (astronomy) iratkozott be (Whitney, 1978).

[11] A legtöbben valóban ilyen kicsinynek tartották csillagvárosunkat, bár Kapteyn saját csillagszámálás eredményei alapján Galaxisunkat 50-55 ezer fényév nagyságúnak gondolta (Tóth, 2013).

[12] Shapley szerint Világunk (a Tejútrendszer) 300 ezer fényév nagyságú. A XX. század második felében a 100 maximum 150 ezer fényéves értéket fogadták el. A XXI. század elején azonban már 200 ezer fényévnél is nagyobbnak gondolták galaxisunkat. A legújabb eredmények szerint a Galaxis mérete még a Shapley szerinti 300 ezer fényéves nagyságot is eléri (Fukushima és mtsai, 2019).

**A Nagy Vita részletei**

Eredetileg két 45 perces előadást terveztek, de Shapley szerint 35 perc is elegendő lenne. Curtis kevesellte a 35 percet, így alakult ki a 40 perces előadások kompromisszuma (Hoskin, 1976). Az előadások 1920. április 26-án hangzottak el Washingtonban. Virginia Trimble (1943- ) a vita 75 éves évfordulóján közöl egy összefoglaló cikket (Trimble, 1995), melyben említi a 75 éves évfordulón a gammakitörésekről tartott vitát is. Ezekről és még két másik csillagászat témájú vitáról tájékozódhat az olvasó a NASA Great Debates in Astronomy oldalain,[13] melyen megtalálható a Shapley - Curtis vita publikált anyaga (Shapley és Curtis, 1921), az azzal kapcsolatos publikációk, valamint a két résztvevő nekrológja. Trimble az említett cikkében pontokba szedte a vitán elhangzott állításokat. Ezek a pontok a következők:

1, A gömbhalmazokban megfigyelt F, G és K csillagokról Shapley azt állította, hogy F, G, K óriás csillagok -3 körüli abszolút magnitúdóval. Ennek következtében a gömbhalmazok átlagos távolsága 10-30 kpc. Curtis azt gondolta, hogy ezek a csillagok F, G, K törpék nagyjából +7 abszolút magnitúdóval, tehát a gömbhalmazok átlagos távolsága 1-2 kpc.

Amikor a Palomar hegyi 5 méteres Hale távcsővel történt megfigyelésekkel a szín-fényesség diagrammon már meg tudták figyelni a főserozati töréspontot (lásd például Sandage 1953), kiderült, hogy alapvetően Shapleynak volt igaza.

2, A gömbhalmazokban megfigyelt B típusú csillagokról Shapley azt állította, hogy 0 körüli abszolút magnitúdóval rendelkeznek, hasonlóan a Nap közeli késői-B és korai-A csillagokhoz. Curtis azt állította, hogy valami probléma van az adatokkal, mert a Naprendszerhez közeli legfényesebb kék csillagok fényesebbek, mint a legfényesebb vörös csillagok, míg a gömbhalmazokban ez fordítva van.

Walter Baade az 1940-es évek elején oldotta meg ezt a kérdést a kétféle csillagpopuláció felfedezésével. Tehát mind a két félnek igaza volt a saját állításában.

3, A cefeidák mint távolságindikátorok. Shapley a Nagy Magellán-felhőben talált periódus-luminozitás relációt használta a gömbhalmazok távolságának a meghatározásához. A null-pont kalibrálásához néhány a galaktikus síkban levő ismert távolságú cefeidát használt. Curtis szerint a galaktikus cefeidáknál nem bizonyított a periódus-luminozitás reláció. Több adat kell, állította.

A megoldást ismét Baade szolgáltatta az 5 méteres távcsővel, mivel nem talált RR Lyrae változókat az Androméda-galaxisban. Curtisnek igaza volt, hogy több adatra van szükség, de a több adat Shapleyt igazolta, a cefeidák jó távolságindikátorok, de tudni kell, hogy több fajta cefeida van (legalább háromféle, I-es típusú, II-es típusú és RR Lyrae).

4, Jól használható módszer-e a spektroszkópiai parallaxis. Shapley szerint igen, ha az óriások felszínén uralkodó gravitációra utaló vonalarányokat meg tudjuk figyelni a közeli csillagokban. Curtis szerint ezek csak a kalibráláshoz használt csillagok 100 parszeken belüli régióiban használhatók.

Shapleynek ebben is igaza volt.

5, Curtis a csillagszámlálásokra hivatkozva állította, hogy a Tejútrendszer (relatíve) kicsi. Curtis szerint a spirál-ködökben megfigyelt por csak a csillagkorongon kívül található, ezért a csillagok között nincs jelentős por. Shapley erről a témáról nem beszélt, talán azért, mert ő gömbhalmazokkal kapcsolatos elemzései során szintén elhanyagolta a csillagközi elnyelést.

---

[13]  A Jerry Bonnell és Robert Nemiroff által kezelt https://apod.nasa.gov/diamond_jubilee/debate.html oldalon összefoglaló található az 1996-os Tammann (1932-2019) és van den Bergh (1929- ) közti (The Scale of the Universe), valamint az 1998-as Peebles (1935- ) és Turner (1949- ) (The Nature of the Universe) közti vitáról. A 75 éves évfordulón 1995-ben tartott vita (The Distance Scale to Gamma-ray Bursts) a gammakitörések eredetéről folyt. Bohdan Paczynski (1940-2007) képviselte a kozmikus eredetet, míg Donald Q. Lamb (1945- ) a kitörések galaktikus eredete mellett érvelt.

Ebben mind a ketten tévedtek, mint az Trumpler méréseiből később kiderült (Trumpler, 1930) a galaktikus síkban a por és ez által az elnyelés jelentős, és így fontos tényező.

6, Csillagfejlődés. Ennél a pontnál Wirginia Trimble megjegyzi (Trimble, 1995), hogy abban az időben a Naprendszer keletkezésére nem a ma elfogadott gázködökből való kialakulás, hanem a Jeans féle csillagok ütközése általi kialakulás volt elfogadott. Itt újra érdemes felhívni a figyelmet, hogy 1920-ban nem csak a csillagok kialakulását és működését nem értettük még, hanem felfedezésre várt lényegében a teljes kvantummechanika, Schrödinger- és Dirac-egyenlet, Heisenberg-féle határozatlansági reláció, valamint nem csak a pozitron, hanem a neutron is, így az atommag összetétele sem volt ismert, hogy a magerőket és a magenergiát már ne is említsük. Trimble mindazonáltal e pontban Curtis véleményét tartja helyesnek, aki azt mondta, hogy a spirálködöknek nincs köze a csillagkeletkezéshez, ami valóban igaz.

7, A spirál-ködök égi eloszlása. Mint köztudott a Tejút síkjában alig-alig látunk spirál-ködöket, míg a galaktikus pólusok környezetében hemzsegnek. Simon Singh állításával ellentétben Shapley nem beszélt erről a témáról (Singh, 2004). Curtis szerint semmi nem zárja ki, hogy úgy, mint más spirál-ködök esetében a galaktikus lemezen kívül egy porgyűrű legyen, ami megakadályozza a távoli halvány ködök megfigyelését.

Curtis nem járt nagyon messze a valóságtól, de elvétette a lényeget, hogy a porködök a csillagokkal összekeveredve a Tejútrendszer részei.

8, Nóvák abszolút fényessége. Mindketten egyetértettek abban, hogy nóva, „új csillag" (*nova stella*) figyelhető meg a Tejútban és a spirál-ködökben. Shapley szerint az a feltételezés, hogy a spirálisok galaxisok, elfogadhatatlanul nagy fényességet eredményezne e csillagoknak. Curtis szerint, azonban a Tejútrendszer sokkal kisebb, így a spirál-ködök is, vagyis nincs gond a közelebbi, de Tejúton kívüli hasonló méretű galaxisokkal és a nóvákkal. Azt Curtis is elismerte, hogy az 1885-ben megfigyelt S Andromedae csillag fényesebb, mint az átlag nóva, illetve a Tycho féle 1572-ben megfigyelt is hasonló lehetett. Szerinte nem kizárt, hogy kétféle nóva típus létezzen. Helytelen feltételezésből (a Tejút kis mérete) helyes sejtésre jutott.

1933-ban Baade és Zwicky alkotta meg a szupernóva szót és egy új osztályát a nóváknak. Megjegyezzük, hogy a nóvák kalibrációjánál négy megfigyelt esemény volt. Ezt a kalibrálást Curtis hajlandó volt elfogadni, míg hasonló kalibrálást a galaktikus cefeidák esetében nem.

9, A Slipher által mért nagy sebességekre a spirál-ködöknél, egyikük se tudott helyes magyarázatot adni. Hubblenek és Humasonnak az 1920-as évek végén tett megfigyelései nyomán sikerült a kérdést tisztázni (táguló Világegyetem).

10, Shapley rámutatott, hogy a spirál-ködök központjának a fényessége sokkal fényesebb, mint ahogy azt a Tejútban mérjük. Curtis e témában csendben maradt. Shapleynek igaza volt a mért adatokban, de a magyarázata helytelen volt. A magyarázat itt is a csillagközi anyag elnyelésében rejlett.

11, Curtis mérésekre hivatkozva azt állította, hogy spirálisok színe és spektrum vonalai csillagszerűek, illetve a csillaghalmazokéhoz hasonló, ezért maguk is csillagok halmazai.

E témában Shapley maradt hallgatag. Mint tudjuk Curtisnek igaza volt.

12, A Napunk a galaxis centrumában van? Shapley szerint nem. Szerinte a ma Gould-övnek nevezett csillagcsoportosulás becsapta a többieket. Curtis szerint a központban vagyunk.

E témában Shapleynek volt igaza.

13, A spirálisok forgása. Van Maanen mérései szerint a spirál-ködök forognak (Maanen, 1916). Shapley szerint ez végzetes csapás a galaxis teóriának. Curtis egyetértett, de szerinte a mérés nem meggyőző, mivel a hiba megközelíti a mért értéket.

Curtisnek lett igaza. Van Maanen jó barátja volt Shapleynek, ezért Shapley könnyen hitt neki. Viszont a forgás kimutatása valóban perdöntő lett volna.

**Mindketten győztek, és egyikük se**

A fentiekből is látszik, hogy nehéz eldönteni kinek volt igaza, de talán nem is szükséges. Az olvasó, ha akarja maga is számolhat egy eredményt a fenti pontok alapján. Ez esetben is döntetlen közeli eredmény adódik. A legtöbb valóban szakmai írás, úgy fogalmaz, hogy mindketten győztek, és egyik se győzött. Amikor saját megfigyeléseikre támaszkodtak, akkor igazuk volt (pl. Shapley fotometriai megfigyelései a gömbhalmazok csillagairól, vagy Curtis spirálköd megfigyelési), de problémás volt, ha mások által megfigyelt adatokra támaszkodtak (pl. van Maanen megfigyelése a spirálisok forgásáról). A legtöbb szakirodalom megemlíti, hogy Shapley közelebb járt az igazsághoz a Tejútrendszer méretével kapcsolatban, valamint a Nap helyzetéről a Galaxisban. Curtisnek igaza volt a spirálködökkel kapcsolatban, és jól sejtette meg, hogy két fajta nova is létezhet. De mindketten tévedtek több dologban (pl. a csillagközi por elhanyagolhatóságát illetően).

Sajnos a legelterjedtebb nézet, hogy a vita a spirál galaxisokról szólt, és így a vitát Curtis nyerte. A YouTube-on több videó is található erről. Maga a vita lezajlása is érdekes, ugyanis nem volt vita, hanem két előadás. Az előadások előtt hosszabb díjátadási ceremóniát tartottak, így a közönség már elég fáradt volt. Több forrás is helytelenül állítja, hogy Einstein jelen lett volna (például Ferris, 1985, Singh, 2004, Tóth, 2013).[14] Az Einstein életrajzokban ez leellenőrizhető, lévén Einstein első amerikai útja 1921-ben volt, nem lehetett 1920-ban az előadásokon jelen. Egy másik többször előforduló tévedés, hogy a vita 1921-ben lett volna. Ennek alapja, hogy a két előadás nyomtatott formában csak 1921-ben jelent meg (Shapley és Curtis, 1921). Shapley figyelembe vette, hogy a hallgatóság többsége nem volt szakcsillagász, ezért a legegyszerűbb fogalmakat is részletesen elmagyarázta. Curtis viszont szakmaibb előadást tartott, ezért a helyszíni benyomás az lehetett, hogy Curtis volt a szakszerűbb. Curtis nagy súlyt helyezett a spirálisokra, míg Shapley az előadás címére koncentrált, mekkora az Univerzum. Shapley így emlékszik vissza később:

„Témánk a Világegyetem léptéke volt. Beszédemben erre készültem, és erről szóltam. Én úgy hiszem, megnyertem a vitát a megjelölt témában. Nekem volt igazam, és Curtis tévedett a legfontosabb dologban, a méretben. A Világegyetem nagy, ő pedig kicsinek gondolta. Curtis kezdettől fogva más témáról beszélt. Arról, hogy a spirálgalaxisok a mi rendszerünkben vagy rajta kívül vannak? Szerinte kívül vannak. Én pedig azt mondtam, hogy én nem tudom mik azok, de a meglévő bizonyítékok szerint nem kívül vannak. De nem ez volt a megjelölt téma. Curtis szalmabábbal hadakozott és azt legyőzte." (Whitney, 1978) A Világegyetem ebben a szövegben akkor a Tejútrendszert jelentette.

Ezzel ellentétben, mint azt Ferris is leírja (Ferris, 1977) az a nézet terjedt el, hogy Curtis legyőzte Shapleyt. Megismételjük, hogy egyes állításokban egyiküket igazolták a későbbi kutatások, másokban a másikukat. Valóban a két legfontosabb állítás, a spirálködök mibenléte, és a Tejútrendszer mérete. Az előbbiben Curtisnek az utóbbiban Shapleynek lett igaza. Shapleynek abban igaza volt, hogy az előadások címe is a Világunk mérete volt, és bár Shapley kissé túlbecsülte a Tejútrendszer méretét, mégis közelebb állt a valósághoz, mint Curtis. Nem beszélve az abban elfoglalt helyzetünkről. Curtis valóban többet beszélt a spirálisokról, és utólag is ez vonzott nagyobb érdeklődést. Ha e két fontos témát vesszük (szándékosan nem számítjuk a Naprendszer helyzetét a galaxisban), akkor is döntetlenre áll a vita. Valóban igazságtalan Shapley-vel, hogy a köztudatban a spirálgalaxisok mibenlétét tekintik a vita egyetlen említésre méltó tárgyának. Milyen volt a vita idején a tudományos közvélekedés? Túlnyomó többség a kisebb méretű Galaxis nézetet fogadta el, és a többség abban is Curtis-szel értett egyet, hogy a spirál-ködök nem a Galaxisban találhatók. Éppen ezért Curtis helyett más is érvelhetett volna, de talán senki nem helyettesíthette volna Shapleyt. Shapley teljesen egyedül határozta meg a gömbhalmazok távolságát a cefeidákra támaszkodva, és határozta meg a Galaxis központját, mint távoli helyet. Fontos felhívni még a figyelmet arra, hogy az előadások publikált változata (Shapley és Curtis, 1921) terjedelemben lényegesen meghaladja az előadottakat. Negyven perc bizonyosan nem volt elegendő a nyomtatásban megjelentek részletes kifejtésére.

---

[14] Singh szerint az unalmas felvezetés alatt Einstein odasúgta a szomszédjának „Éppen most gondoltam ki egy új örökkévalóság-elméletet" (Singh, 2004).

Mint írtuk sok részletben egyiküknek, sokban másikuknak volt igaza. Megállapítható, hogy két jól felkészült szakcsillagász tartott egy-egy kiváló előadást, a kor legfrissebb eredményeit összefoglalva. Ezzel kapcsolatban érdemes idézni Shu könyvét (Shu, 1982): „A Shapley-Curtis vita nem csak érdekes történelmi dokumentum. Érdemes tanulmányozni, ahogy két kiváló tudós, hiányos, és néha téves adatokból ellentétes következtetésekre jut." Az 1920-as és részben az 1930-as években a két fő kérdés; a Galaxis mérete (és benne helyünk) és a spirál-ködök mibenléte és mérete nagy viták tárgyát képezte jelentős számú csillagász részvételével. De a „Nagy Vita" kifejezés ezzel a témával kapcsolatban mindig az 1920-as két előadásra utalt. A Nagy Vita szellemi hatását érzékeltetik a már említett további csillagászati nagy viták megtartása a 75. évforduló után. A 100 éves évfordulón érdemes megemlékeznünk arról, hogyan formálta a Nagy Vita és az általa generált további kutatások a Világról alkotott elképzelésünket.

**Hivatkozásjegyzék**